\documentclass[a4paper]{article}

\usepackage{interspeech2013}
\ninept
\usepackage[pass, verbose]{geometry}

\usepackage{tipa}

\usepackage{mathptmx}
\usepackage{ifxetex}
\ifxetex
  \usepackage{fontspec}
\else
  \usepackage[utf8]{inputenc}
\fi

\usepackage{siunitx}
\usepackage{relsize}

\usepackage[inline, shortlabels]{enumitem}

\usepackage{graphicx}
\usepackage{xcolor}
\usepackage{tikz}
\usepackage{adjustbox}

\usepackage{booktabs}

\usepackage{caption}
\usepackage{subfig}

\usepackage{listings}
\lstdefinelanguage{bvh}{keywords=[1]{HIERARCHY, ROOT, OFFSET, CHANNELS, End, Site, MOTION, Frames, Frame, Time},
  keywords=[2]{XPosition, YPosition, ZPosition, XRotation, YRotation, ZRotation},
  keywords=[3]{T3, upperlip}}

\usepackage{cite, IEEEtrantools}
\usepackage{doi}

\usepackage{hyperref}
\hypersetup{colorlinks, linkcolor=black, urlcolor=black, citecolor=black}
\usepackage[all]{hypcap}

\usepackage[acronym, shortcuts, nowarn]{glossaries}
\glsdisablehyper
\newacronym{2d}{2D}{two-dimensional}
\newacronym{3d}{3D}{three-dimensional}
\newacronym{ascii}{ASCII}{American Standard Code for Information Interchange}
\newacronym{av}{AV}{audiovisual}
\newacronym{bvh}{BVH}{Biovision Hierarchy}
\newacronym[longplural=degrees of freedom]{dof}{DOF}{degree of freedom}
\newacronym{ema}{EMA}{electromagnetic articulography}
\newacronym{EMA}{\acs{ema}}{\acl{ema}}
\newacronym{est}{EST}{Edinburgh Speech Tools}
\newacronym{fem}{FEM}{finite element modeling}
\newacronym{ik}{IK}{inverse kinematics}
\newacronym{gui}{GUI}{graphical user interface}
\newacronym{mri}{MRI}{magnetic resonance imaging}
\newacronym{nurbs}{NURBS}{non-uniform rational B-spline}
\newacronym{rmse}{RMSE}{root-mean-square error}
\newacronym[longplural=regions of interest]{roi}{ROI}{region of interest}
\newacronym{uti}{UTI}{ultrasound tongue imaging}
\newacronym{vcv}{VCV}{vowel-consonant-vowel}
\makeglossaries

\newcommand{\keywords}{%
  speech production,
  articulatory data,
  \acl{EMA},
  vocal tract,
  motion capture,
  visualization
}
\newcommand{\mngu}{{\tt mngu0}}

\usepackage{authoraftertitle}
\title{Speech animation using \acl{EMA}\\
as motion capture data}

\makeatletter
\def\name#1{\gdef\@name{#1\\}}
\makeatother
\name{%
  {\em Ingmar Steiner}\,$^\textnormal{1,2}$,
  {\em Korin Richmond}\,$^3$,
  {\em Slim Ouni}\,$^4$
}

\address{%
  $^1$Multimodal Computing and Interaction, Saarland University, Germany; \\
  $^2$DFKI GmbH, Saarbrücken, Germany; \\
  $^3$CSTR, University of Edinburgh, UK; \\
  $^4$Université de Lorraine, LORIA, UMR 7503, France \\
  {\small
    \nolinkurl{ingmar.steiner@dfki.de};
    \nolinkurl{korin@cstr.ed.ac.uk};
    \nolinkurl{slim.ouni@loria.fr}
  }
}

\hypersetup{%
  pdfinfo={%
    Title={\MyTitle},
    Author={Ingmar Steiner, Korin Richmond, Slim Ouni},
    Keywords={\keywords},
    Subject={AVSP 2013, Annecy, France}
  }
}

\begin{document}

% apply IEEE BibTeX customizations
\bstctlcite{IEEEexample:BSTcontrol}

% title
\maketitle

% abstract
\glsresetall
\glsunset{2d}
\glsunset{3d}
\glsunset{ascii}
\begin{abstract}
  \Ac{ema} captures the position and orientation of a number of markers, attached to the articulators, during speech.
  As such, it performs the same function for speech that conventional motion capture does for full-body movements acquired with optical modalities, a long-time staple technique of the animation industry.
  
  In this paper, \ac{ema} data is processed from a motion-capture perspective and applied to the visualization of an existing multimodal corpus of articulatory data, creating a kinematic \ac{3d} model of the tongue and teeth by adapting a conventional motion capture based animation paradigm.
  This is accomplished using off-the-shelf, open-source software.
  Such an animated model can then be easily integrated into multimedia applications as a digital asset, allowing the analysis of speech production in an intuitive and accessible manner.
  
  The processing of the \ac{ema} data, its co-registration with \ac{3d} data from vocal tract \ac{mri} and dental scans, and the modeling workflow are presented in detail, and several issues discussed.
\end{abstract}
\glsresetall
\glsunset{2d}
\glsunset{3d}
\glsunset{ascii}

% intro
\section{Introduction}

% adapted from IAST 2012 paper:

Speech scientists have a number of medical imaging modalities at their disposal to capture hidden articulatory motion during speech, including realtime \ac{mri}, \ac{uti}, and \ac{ema} \cite{Hiiemae2003Tongue-Movement}, the latter more recently in \ac{3d} \cite{Hoole2010Five-dimensional}.
Such techniques are commonly used to analyze and visualize the articulatory motion of human speakers.
Indeed, the resulting data has been applied to articulatory animation for \ac{av} speech synthesis \cite{Cohen1993Modeling, Pelachaud1994Modeling, King20013D-parametric, Badin2002Three-dimension, Engwall2003Combining, Fagel2004An-articulation};
using motion-capture data to animate such models can lead to significant improvements over rule-based animation \cite{Engwall2009Real-vs.-rule}.
However, these synthesizers are generally focused towards clinical applications such as speech therapy or biomechanical simulation.

While the lip movements can be acquired using optical tracking and the teeth and jaw are rigid bodies, the tongue is more complex to model, since its anatomical structure makes it highly flexible and deformable.
The majority of previous work has modeled the tongue based on static shapes (obtained from \ac{mri}) and statistical parametric approaches to deforming them by vertex animation \cite{Badin2002Three-dimension, Engwall2003Combining} or \ac{fem} \cite{Stone2001Modeling, Gerard2006A-3D-dynamical, Dang2004Construction, Vogt2007Efficient, Yang2012Physics-Based}.
% end adaptation
\cite{Pelachaud1994Modeling} and \cite{Ilie2012Efficient} are an exception in that they deform the tongue model using a skeletal animation technique, which rigs the tongue mesh on an armature of pseudo-bones exerting varying degrees of influence on the position of the mesh vertices.

In this paper, we describe a technique for articulatory animation, i.e., visualization of movements of the articulators during speech, by adapting a conventional motion capture based animation paradigm (\autoref{sec:mocap}).
%, by deforming a static \ac{3d} model of the tongue and teeth exclusively by \ac{ema} data.
A form of skeletal animation is applied to the tongue, but driven directly by \ac{ema} data, without the intermediate abstraction of independent parameters.
The static \ac{3d} model is extracted from volumetric \ac{mri} and dental scans from the same speaker as the \ac{ema} data.
In this way, we avoid the issue of cross-speaker vocal tract normalization.

The reader is kindly requested to note that this technique is by no means intended to provide an accurate model of tongue shapes or movements, as previous work using biomechanical models does;
rather, the advantage here is the lightweight implementation, which relies exclusively on the articulatory data itself, processed as described in \autoref{sec:processing}, and a conventional \ac{3d} modeling workflow using industry standards and off-the-shelf, open-source software (cf.\ \autoref{sec:modeling});
this eliminates the implementation issues which burden more ambitious frameworks \cite[for instance]{Vogt2007Efficient}.

\section{\acs{ema} as motion capture}
\label{sec:mocap}

Motion capture has been used for many years to analyze human movements, gait and gestures, and to control the motions and expressions of avatars and virtual characters in a variety of media \cite{Menache2000Understanding}.
Accordingly, an entire industry has formed around the acquisition and processing of motion capture data, and the rigging and animation of \ac{3d} models that are controlled using this data.
There is a huge community of producers and consumers of motion capture, countless databases and stock animation resources, and a rich ecosystem of proprietary and open-source software tools to manipulate them.

Despite attempts to design and promote elaborate and flexible universal standards for motion capture data \cite{c3d, Chung2004MCML}, certain other industry-backed file formats have evolved to become de-facto standards instead \cite{MeredithMotion-Capture}.
One format in particular, \ac{bvh}, has survived the company that created it and is now widely supported, presumably because it is simple and clearly defined, straightforward to implement, and human-readable \cite{GleicherBiovision-BVH, HowardBVH}.

A \ac{bvh} file contains \ac{ascii} text divided into two sections, which define the skeletal objects underlying the motion capture data in a \texttt{HIERARCHY} of joints, and the \texttt{MOTION} of these objects, respectively.
Although most \ac{bvh} data describes the shape and motion of bipedal humanoids, this is by no means a requirement;
multiple independent objects of arbitrary internal structure can be defined.

This feature of the \ac{bvh} format, along with its widespread support, is a simple, but fundamental point for the premise of the present paper.
By interpreting the coils of \ac{ema} data as objects whose motion is captured over time and describing it in a format such as \ac{bvh}, it becomes straightforward to introduce the \ac{ema} data into a conventional motion capture processing workflow and to use it to control the visualization of geometric models of speech articulation.

\section{Articulatory animation}

Based on previous work exploring the feasibility of articulatory animation based on \ac{ema} in a motion capture paradigm \cite{Steiner2012Artimate}, we apply our articulatory modeling approach to a large multimodal corpus of articulatory data.

While we had previously experimented with a layout of seven \ac{ema} coils on the tongue, we found that the resulting animation was vulnerable to noise and errors in the seven-coil data, potentially due to the influence of any or all of the following factors:%
\begin{enumerate*}[label={~(\alph*)}, itemjoin={,}, itemjoin*={,~and}]
  \item coil detachments during the recording session
  \item faulty coil hardware
  \item interference due to coil proximity
  \item issues with the post-processing software \cite{Stella2012Numerical}
\end{enumerate*}.

\subsection{Multimodal corpus}

The \mngu\ corpus \cite{mngu0} contains articulatory data obtained from a single male speaker of British English, using a variety of modalities.
This data is freely available for research purposes.
Several hours of speech were recorded using a Carstens AG500 articulograph and video camera, yielding \ac{3d} \ac{ema} data and synchronized video and audio for more than \num{2000} utterances  \cite{Richmond2011Announcing}.
In addition, a set of volumetric and dynamic \ac{mri} scans were acquired of the speaker's vocal tract during sustained and \ac{vcv} speech production, respectively \cite{Steiner2012Magnetic}.
Finally, dental casts were made of the speaker's teeth, gums, and palate;
the casts were then scanned to produce high-resolution, digital \ac{3d} models.

\subsection{Data processing}
\label{sec:processing}

\subsubsection{\Ac{ema} data}

\begin{lstlisting}[float=t,
  caption={%
    Excerpt of one \ac{bvh} file (\texttt{mngu0\_s1\_0001.bvh}).
    Note that only two \ac{ema} coil objects (\emph{T3} and \emph{upperlip}) are listed, and the motion data (shown for all 10 coils) is truncated after the second frame
  },
  label=lst:bvh, linerange={1-1,74-96}, language=bvh]
HIERARCHY
ROOT T3
{
	OFFSET	-0.083	-0.686	0.722
	CHANNELS 6 XPosition YPosition ZPosition XRotation YRotation ZRotation
	End Site
	{
		OFFSET	0.000	-1.000	0.000
	}
}
ROOT upperlip
{
	OFFSET	-0.945	0.297	-0.130
	CHANNELS 6 XPosition YPosition ZPosition XRotation YRotation ZRotation
	End Site
	{
		OFFSET	0.000	-1.000	0.000
	}
}
MOTION
Frames:	724
Frame Time:	0.005
6.790	12.273	1.200	-8.137	56.687	-1.782	-6.083	12.965	0.386	19.706	-11.455	-52.567	0.175	0.978	-3.002	3.939	24.401	-51.690	0.053	-0.487	-2.579	56.493	9.144	-2.778	0.005	-0.006	5.717	56.740	1.991	7.709	-0.003	-0.026	0.006	-57.279	-1.344	0.390	0.258	1.551	-0.986	7.964	-30.909	-47.582	0.106	3.215	0.254	-8.542	-56.293	-6.396	0.112	5.056	-0.049	-4.804	-39.311	41.405	0.096	-1.061	-0.194	-54.183	17.068	-7.465
6.789	12.272	1.201	-8.140	56.686	-1.790	-6.083	12.966	0.386	19.704	-11.460	-52.566	0.172	0.979	-3.003	3.946	24.397	-51.692	0.050	-0.489	-2.579	56.493	9.144	-2.788	0.005	-0.006	5.717	56.741	1.995	7.702	-0.003	-0.025	0.006	-57.279	-1.347	0.390	0.260	1.550	-0.986	7.962	-30.937	-47.564	0.107	3.215	0.254	-8.544	-56.295	-6.380	0.114	5.058	-0.051	-4.791	-39.304	41.413	0.094	-1.057	-0.196	-54.183	17.063	-7.476
\end{lstlisting}

For this paper, the ``day 1'' subset of the \ac{ema} data in the \mngu\ corpus was used;
the layout features \ac{ema} coils on the upper and lower lip, three tongue coils (T1-3, on the tip, blade, and dorsum, respectively), and mandibular incisors, as well as reference coils behind each ear, on the bridge of the nose, and on the maxillary incisors.

The large number of utterances and unusually high quality of the data greatly outweigh the sparse coverage of the tongue surface represented by only three coils.
Nevertheless, it turns out that the spline \acs{ik} based animation (cf.\ \autoref{sec:rigging}) yields satisfactory results with this arrangement.

The \mngu\ \ac{ema} data is provided in binary \texttt{EST\_Track} format, which can be manipulated using the \ac{est} \cite{est}.
The data was processed directly from the raw \texttt{amp} files produced by the AG500, using a custom version of TAPADM \cite{tapadm}.
The \ac{ema} coil orientation is encoded not as two Euler angles, as in the \texttt{pos} file format produced by the Carstens software \cite{ag500}, but as three rotation normals.

The \texttt{EST\_Track} files were converted to \ac{bvh} format in the following manner.
Each \ac{ema} coil is interpreted as an independent armature in the \texttt{HIERARCHY} and encoded as a separate \texttt{ROOT} object with six channels corresponding to the Cartesian coordinates and rotation normals of the coil's position and orientation, respectively.
The values for each coil's \texttt{OFFSET} from the origin represent the coil distribution in the first data frame.
Each armature is terminated by a tip whose \texttt{OFFSET} is given by a unit vector.
The \texttt{MOTION} is simply copied from the corresponding channels in the \texttt{EST\_Track} data.
An example of the result of this conversion is given in \autoref{lst:bvh}.

For downstream registration using the palate as a landmark (cf.\ \autoref{sec:registration}), in lieu of a dedicated palate trace sweep, the convex hull of the tongue coil position samples was calculated as a contour intersected with the speaker's midsagittal plane, using VisArtico \cite{visartico}.
This reconstructed palate contour was then exported as a \ac{3d} point cloud.

\begin{figure*}
  \subfloat[Tessellated voxels from \ac{mri} scan]{%
    \includegraphics[trim={4cm 0 3cm 0}, clip, width=0.3\textwidth]{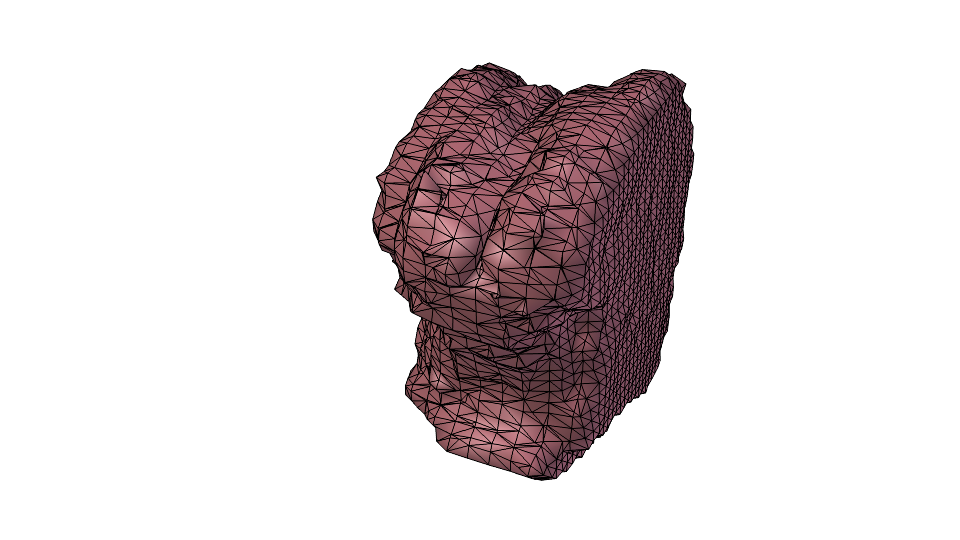}
    \label{fig:retopo_before}
  }
  \hfill
  \subfloat[Retopology cage]{%
    \includegraphics[trim={4cm 0 3cm 0}, clip, width=0.3\textwidth]{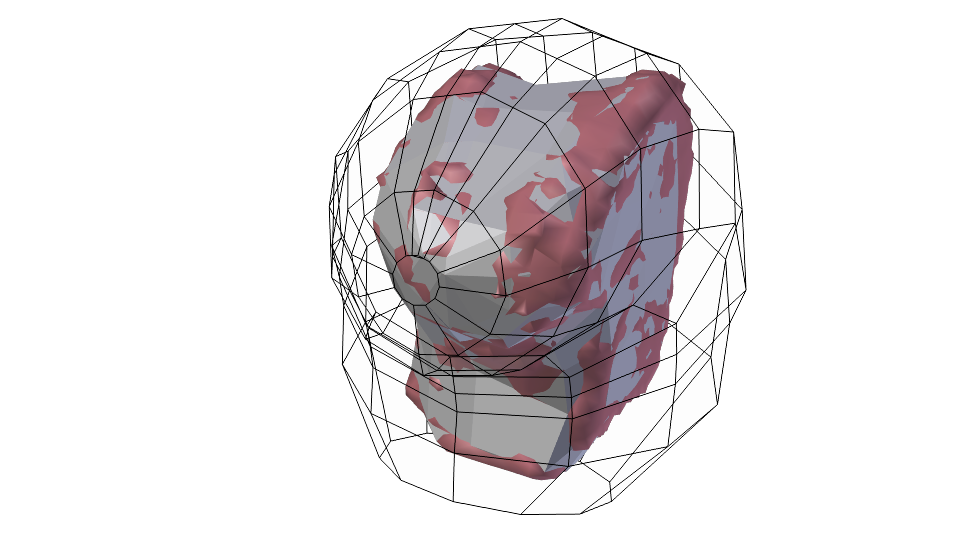}
    \label{fig:retopo_during}
  }
  \hfill
  \subfloat[Smoothed mesh used for deformation]{%
    \includegraphics[trim={4cm 0 3cm 0}, clip, width=0.3\textwidth]{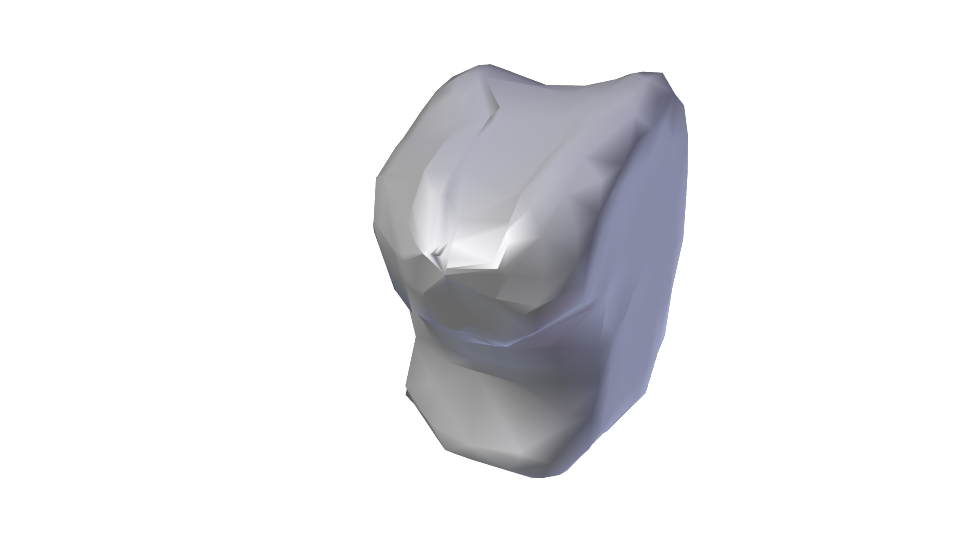}
    \label{fig:retopo_after}
  }
  \caption{Tongue mesh retopology}
  \label{fig:retopology}
\end{figure*}

\subsubsection{Tongue \ac{mri} data}

Instead of an artificial static \ac{3d} model of the tongue and teeth, as was used for our feasibility study \cite{Steiner2012Artimate}, the present paper uses a static tongue mesh extracted from the \ac{mri} subset of the \mngu\ corpus.
Selecting an articulatory configuration with a clearly visible cavity between the tongue and palate surfaces, the tongue from the volumetric \ac{mri} scan of the sustained vowel \textipa{[A:]} was manually segmented and exported, using OsiriX \cite{osirix}, which acts as a \ac{gui} to a number of toolkits for manipulating and visualizing medical imaging data.

The tongue was first enclosed in a \ac{roi} annotation in each \ac{mri} slice, drawn by hand using a pen tablet.
The pixel values outside these \acp{roi} were then set to zero.
Finally, the tongue was rendered as an isosurface within the \ac{3d} \ac{roi} formed by interpolating between the slicewise annotations, using a threshold value, and exported as a static \ac{3d} mesh (\autoref{fig:retopo_before}).

Since this mesh is generated simply by tessellating the anamorphic voxels of the volumetric \ac{mri} data, the mesh topology is extremely ill-suited to the goal of realistic deformation.
As a consequence, the tongue mesh was retopologized with a simple cage, which was ``shrinkwrapped'' to the surface of the exported mesh (\autoref{fig:retopo_during}) and smoothed using Catmull-Clark subdivision \cite{Catmull1978Recursively} (\autoref{fig:retopo_after});
this task was performed using Blender \cite{blender}.

In order to obtain a landmark to be used in subsequent registration (cf.\ \autoref{sec:registration}), the hard palate was also segmented and exported as a static mesh in a manner analogous to the tongue extraction.

\subsubsection{Dental scans}

%\begin{figure}
%  \includegraphics[width=\columnwidth]{teeth_mesh}
%  \caption{Dental meshes after vertex decimation}
%  \label{fig:dental}
%\end{figure}

As the resolution of the dental scans was deemed too high for efficient processing, the vertices of the maxilla and mandible meshes were deduplicated (using MeshLab \cite{meshlab}) and decimated (using Blender) to roughly \SI{5}{\percent};
this reduced the vertex count from \num{927282} and \num{836892} to \num{7729} and \num{6976} for the maxilla and mandible, respectively.

\subsection{Vocal tract modeling}
\label{sec:modeling}

The \ac{mri} and dental cast scans were co-registered, and the resulting static \ac{3d} model was rigged and animated with the \ac{ema} data, as described below.

\subsubsection{Cross-modal registration}
\label{sec:registration}

\begin{figure}
  \subfloat[\Ac{ema} only;
      the palate contour is shown as a dotted line]{%
    \includegraphics[trim={2cm 2cm 2cm 2cm}, clip, width=\columnwidth]{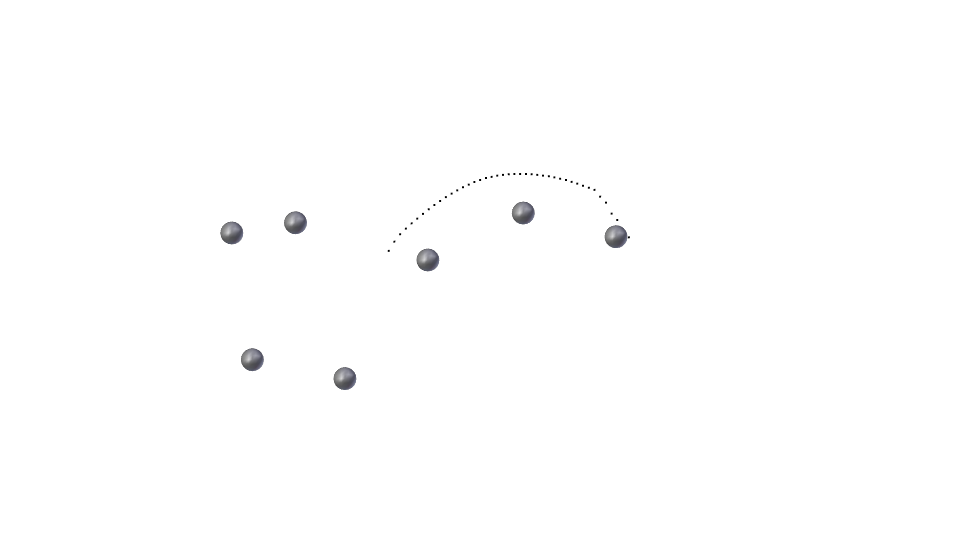}
  }
  \\
  \subfloat[Dental scans added]{%
    \includegraphics[trim={2cm 0 2cm 0}, clip, width=\columnwidth]{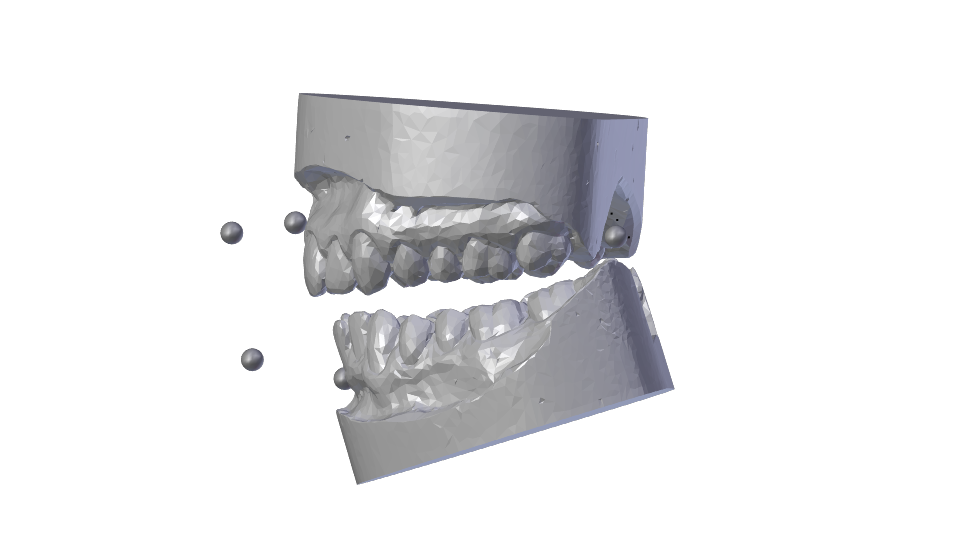}
  }
  \caption{Registration of \acs{ema} data and dental scan meshes}
  \label{fig:registration}
\end{figure}

Using the palate as a landmark, the data from the different modalities was registered into the same geometric space.
The palate contour from the \ac{ema} data was used to position the maxillary dental mesh, while the mandibular mesh was initially positioned accordingly, using the occlusal plane as a reference (\autoref{fig:registration}).
The retopologized tongue mesh was co-registered with the other modalities by fitting the surface of the exported palate mesh to the palate surface of the maxillary dental scan.

\begin{figure*}
  \subfloat[Tongue mesh with spline and hooks]{%
    \includegraphics[width=0.3\textwidth]{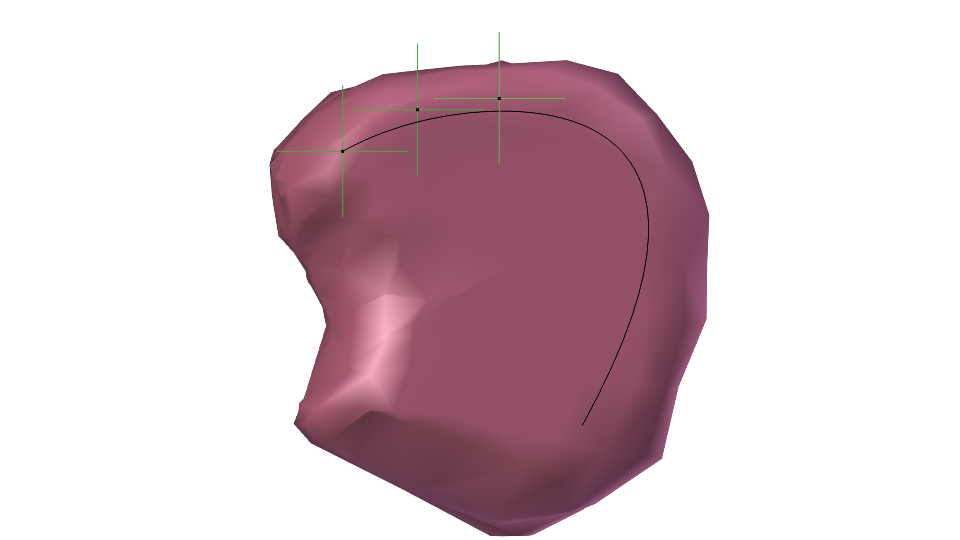}
    \label{fig:hooks}
  }
  \hfill
  \subfloat[Envelopes which determine the vertex deformation weights for each armature joint]{%
    \includegraphics[width=0.3\textwidth]{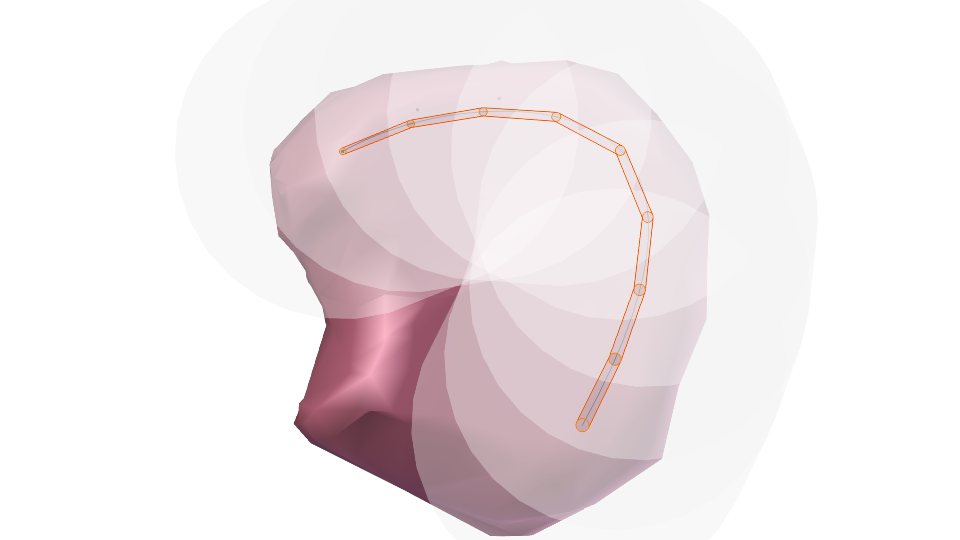}
    \label{fig:envelopes}
  }
  \hfill
  \subfloat[Animated model, with tongue body raised towards a velar constriction.
  The maxilla is hidden to permit a better view of the tongue surface]{%
    \includegraphics[width=0.3\textwidth]{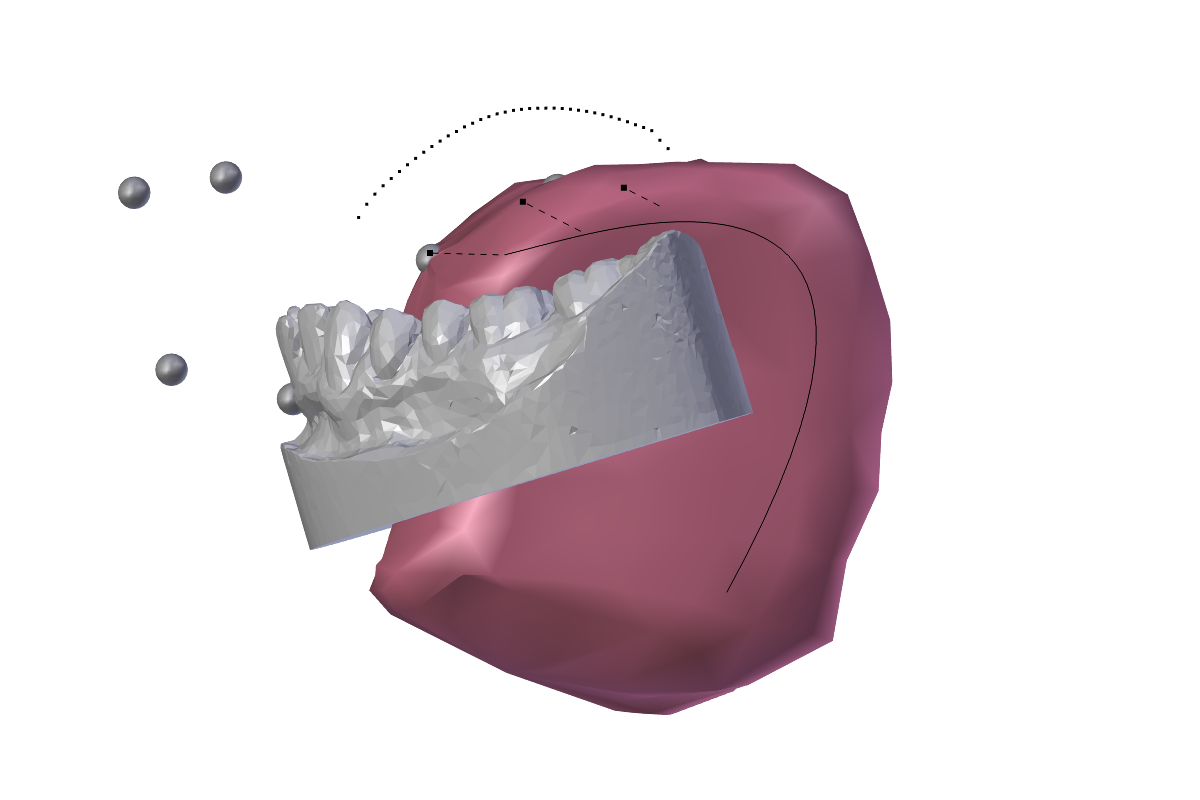}
    \label{fig:animation}
  }
  \label{fig:rigging}
  \caption{Rigging and animating the articulatory model}
\end{figure*}

\subsubsection{Rigging}
\label{sec:rigging}

A \ac{3d} \ac{nurbs} was created along, and just under, the surface of the tongue mesh, from the tongue root to the tip in the midsagittal plane.
The three \ac{ema} tongue coils were then configured to act as ``hooks'', or control points, modifying the shape of the spline according to their deviation from the initial bind pose (\autoref{fig:hooks}).
The hooks were then assigned as children of the respective tongue coils, so that the \ac{ema} data determines their positions.

An armature ``chain'' was then created and parented to the \ac{nurbs}, following its shape using spline \ac{ik}.
The armature envelopes for this chain were expanded to enclose the upper surface and sides of the tongue mesh, and vertex groups were created for the mesh to control the tongue's deformation, with weights assigned automatically, based on the envelopes (\autoref{fig:envelopes}).

For jaw motion, a simple modifier was added to rotate the mandibular dental mesh around a hinge, tracking the \ac{ema} coil on the mandibular incisors.

\subsubsection{Animation}

As a result of the rigging process described above, the \ac{ema} data drives the motion of the jaw in the articulatory model, and likewise controls the tongue by deforming the \ac{nurbs}, which in turn modifies the shape of the tongue armature using spline \ac{ik}.
An example of the animation rendered in this manner can be seen in \autoref{fig:animation} and in the supplementary material for this paper.

\subsection{Evaluation}

\begin{figure}
  \input{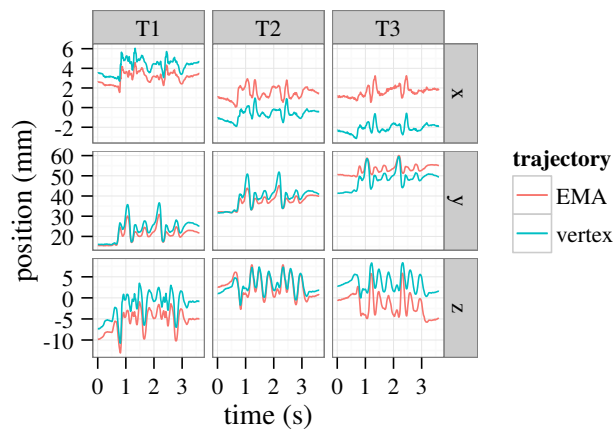}
  \vspace{-\baselineskip}
  \caption{\Acs{ema}-measured and animated trajectories of the three tongue coils and corresponding vertices for one utterance}
  \label{fig:correlation}
\end{figure}

By comparing the motion of the tongue coils with that of three corresponding vertices on the tongue mesh, we receive a rough evaluation of how well the animation preserves the nature of the articulatory data.
While the sparse topology of the mesh does entail a noticeable offset (cf.\ the example in \autoref{fig:correlation}), the overall mean correlation of \num{0.95} indicates that the characteristics of the natural movements are reflected in the animation.

%\begin{table}
%  \centering
%  \begin{tabular}{rrrr}
%    \toprule
%    & T1 & T2 & T3 \\
%    \midrule
%    x & \num{0.98} & \num{0.92} & \num{0.89} \\
%    y & \num{0.99} & \num{0.98} & \num{0.96} \\
%    z & \num{0.99} & \num{0.92} & \num{0.91} \\
%    \bottomrule
%  \end{tabular}
%  \caption{Correlation coefficient for movements of the three tongue coils in the $\{x,y,z\}$ dimensions}
%  \label{tab:correlation}
%\end{table}

It should be noted that the \emph{shape} of the tongue mesh, when it is deformed by the \ac{ema} data for a given articulatory configuration, cannot be expected to match that of static tongue shapes obtained by, e.g., sustained production in a supine posture.
Nevertheless, it could be useful to compare the surfaces of such tongue shapes with those obtained through deformation of the static mesh.
Since corresponding data is available in the \mngu\ corpus, such comparisons are indeed planned in the near future.
A better assessment of the tongue deformation's degree of realism would however be possible by comparing it with real-time \ac{mri} of fluent speech \cite{Narayanan2011A-Multimodal, Niebergall2013Real-time-MRI}

\section{Discussion and Outlook}

While the kinematics of the articulatory animation appear natural, a number of issues were identified with the techniques presented here, and remain to be addressed.

\begin{itemize}[nosep]
  \item The tongue segmentation from \ac{mri} data, and subsequent retopology, represent a rather tedious manual process.
  Automation of some or all of the aspects of these tasks would be highly desirable.
  \item The manual cross-modal registration based on the palate contour is also error-prone.
  Actual \ac{3d} palate trace data in the \ac{ema} modality would permit a more robust surface fitting technique.
  \item There are currently no constraints to prohibit the \ac{3d} models from passing through one another.
  By integrating collision detection using a physics engine such as \cite{bullet}, this could be prevented, and soft body dynamics simulated.
  \item The placement of \ac{ema} coils determines the animation.
  This is both a blessing and a burden, since the lack of independent parameters could cause non-optimal coil layouts to unduly influence the animation.
  \item The initial bind pose in the rigging process is critical to subsequent animation;
  if the exact position of e.g., the jaw coil relative to the mandibular mesh can only be estimated, the overall animation will reflect a poor choice here.
  \item The \ac{mri} data was acquired in a supine position, but the \ac{ema} data was recorded with the speaker sitting upright.
  As a result, the extracted tongue mesh may well be influenced by posture and gravity \cite{Kitamura2005Difference}.
\end{itemize}

In conclusion, we have presented an articulatory animation technique which is driven by articulatory data in a conventional motion capture based animation paradigm.
It features a lightweight implementation using off-the-shelf, open-source software, and a footprint that is small enough to allow integration of the resulting model into applications for articulatory data exploration and real-time visualization, as well as integration into frameworks for \ac{av} speech synthesis for virtual characters, where realistic animation is more important than matching the true shape of the tongue.

Future work includes evaluating the tongue animation based on different tongue shapes or contours in real imaging data, as well as co-registration with the video data in the \mngu\ corpus, based on optical tracking of the visible \ac{ema} coils.

% references
\eightpt
\bibliographystyle{IEEEtran}
\bibliography{paper}

\end{document}